\documentclass{article}
\usepackage{amsmath,amssymb,bbm,amsthm}
\usepackage{fullpage}
\usepackage{thm-restate,color,xcolor,xspace}
\usepackage{hyperref,cleveref}
\usepackage{graphicx}
\usepackage{algorithm,algorithmic}

\usepackage{jason-private}

\begin{document}

\title{Simpler and Faster Directed Low-Diameter Decompositions}
\author{Jason Li\footnote{Carnegie Mellon University. email: \tt jmli@cs.cmu.edu}}
\date{\today}
\maketitle

\begin{abstract}
We present a simpler and faster algorithm for low-diameter decompositions on directed graphs, matching the $O(\log n\log\log n)$ loss factor from Bringmann, Fischer, Haeupler, and Latypov (ICALP 2025) and improving the running time to $O((m+n\log\log n)\log^2n)$.
\end{abstract}

\section{Introduction}
We study the problem of computing a (probabilistic) low-diameter decomposition (LDD) on a directed graph: given a diameter parameter $\De$, delete a random subset of edges such that (1)~each strongly connected component in the remaining graph has weak diameter at most $\De$, and (2)~each edge is deleted with probability at most $\el(n)/\De$ times the weight of the edge, where $\el(n)$ is a loss factor that should be minimized.

Historically, LDDs were first studied on \emph{undirected} graphs, where it has become a central tool in the design of graph algorithms~\cite{linial1993low,bartal1996probabilistic}, especially in constrained computational models such as parallel~\cite{blelloch2011near,miller2013parallel}, distributed~\cite{awerbuch1989network,awerbuch1992fast}, and dynamic~\cite{forster2019dynamic,chechik2020dynamic}. For directed graphs, a non-probabilistic version of LDDs was studied in the context of directed feedback vertex set \cite{seymour1995packing}, minimum multi-cut~\cite{klein1997approximation}, and multi-commodity flow~\cite{leighton1999multicommodity}, though its presence was limited in the following two decades until the seminal result of Bernstein, Nanongkai, and Wulff-Nilsen~\cite{bernstein2025negative}, who used (probabilistic) directed LDDs to obtain the first near-linear time algorithm for negative-weight single-source shortest paths. They devised a simple directed LDD algorithm that achieved a loss factor of $\el(n)=O(\log^2n)$, though optimizing this bound was not their main focus. Directed LDDs were then systematically studied by Bringmann, Fischer, Haeupler, and Latypov~\cite{bringmann2025near}, who improved the bound to $\el(n)=O(\log n\log\log n)$, coming close to the $\Om(\log n)$ lower bound which also holds for undirected graphs. They also designed a randomized algorithm achieving this bound that runs in $O(m\log^5n\log\log n)$ time with high probability,\footnote{\emph{With high probability} means with probability $1-1/n^C$ for arbitrarily large constant $C>0$.} and they remark that one logarithmic factor can be shaved by using a faster single-source shortest paths subroutine.

In this work, we present a simpler and faster algorithm for directed LDDs, achieving the same bound $\el(n)=O(\log n\log\log n)$ and running in $O((m+n\log\log n)\log^2n)$ time, an improvement of two logarithmic factors over the optimized version of~\cite{bringmann2025near}.
\BT\thml{main}
Given a graph with integral and polynomially-bounded edge weights, there is a directed LDD algorithm that achieves loss factor $\el(n)=O(\log n\log\log n)$, runs in $O((m+n\log\log n)\log^2n)$ time, and succeeds with high probability.
\ET

\subsection{Preliminaries}
Let $G=(V,E,w)$ be a directed graph with non-negative and integral edge weights. For a vertex subset $U\s V$, define $G[U]$ as the induced graph on $U$, and define $E[U]$ as the edges of this induced graph. The degree of a vertex $v\in V$ is the number of edges incident to $v$ (which does not depend on the edge weights), and the volume $\vol(U)$ of a vertex subset $U\s V$ equals $\sum_{v\in U}\deg(v)$. For vertices $u,v\in V$, define $d(u,v)$ to be the distance from $u$ to $v$ according to the edge weights. For a vertex subset $U\s V$, define the \emph{weak diameter} of $U$ as $\max\{d(u,v):u,v\in U\}$. For vertex $t\in V$ and non-negative number $r$, define the out-ball $B^+_G(t,r)=\{v\in V:d(t,v)\le r\}$ and in-ball $B^-_G(t,r)=\{v\in V:d(v,t)\le r\}$. We say that an edge $(u,v)$ is \emph{cut} by out-ball $B^+_G(t,r)$ if $u\in B^+_G(t,r)$ and $v\notin B^+_G(t,r)$, and it is \emph{cut} by in-ball $B^-_G(t,r)$ if $v\in B^-_G(t,r)$ and $u\notin B^-_G(t,r)$.

\subsection{Our Techniques}
Our algorithm follows the same template as~\cite{bernstein2025negative,bringmann2025near}: compute an out-ball or in-ball $B$, remove the edges cut by the ball, and recursively solve the induced graphs $G[B]$ and $G[V\sm B]$. Observe that by removing all edges cut by $B$, the vertices in $B$ can no longer share strongly connected components with the vertices in $V\sm B$. Therefore, if the algorithm recursively computes directed LDDs in $G[B]$ and $G[V\sm B]$, which means deleting edges such that each strongly connected component has weak diameter at most $\De$, then deleting these same edges, along with the edges cut by the ball $B$, also produces strongly connected components with weak diameter at most $\De$ in the original graph $G$.

To bound the recursion depth and speed up the running time, the algorithms actually compute multiple balls in each recursive instance so that each recursive component has sufficiently fewer edges.

The algorithm of~\cite{bringmann2025near} adopts a specialized recursive structure colloquially known as Seymour's trick~\cite{seymour1995packing}, and we use the same general approach. At a high level, our algorithm has two main differences which constitute the conceptual contributions of this paper.

\para{CKR instead of geometric ball-growing.}
The algorithm of~\cite{bringmann2025near} uses the geometric ball-growing technique: sample a random radius from a geometric distribution, and cut the appropriate out-ball or in-ball. The main benefit of the geometric distribution is its \emph{memoryless} property, which is crucial in bounding edge cut probabilities. On the other hand, to avoid losing another factor of $O(\log n)$ in the bound of $\el(n)$, the parameters are set in a way that the probability that the radius grows too large is larger than $1/\poly(n)$, so it cannot be ignored with high probability and forces a delicate probability analysis.

Instead, we adopt an approach first used by Calinescu, Karloff, and Rabani~\cite{cualinescu2000improved} in the context of LP rounding for multi-way cut. Here, the radius is no longer geometric, but uniformly random in a given range, and each ball uses the same sampled radius. The twist is that CKR computes the balls in a \emph{random} order, which is key to their analysis.

Similar to \cite{bringmann2025near}, we cannot afford to grow balls from all (remaining) vertices, which would incur a factor of $O(\log n)$ in the bound of $\el(n)$, so we also \emph{sample} a subset of vertices from which to grow balls. On each application of CKR, for a given edge $(u,v)$, the cut probability has a factor that is logarithmic in the number of sampled vertices whose balls can possibly cut $(u,v)$. This factor is carefully balanced with the sizes of the recursive instances by an application of Seymour's trick~\cite{seymour1995packing}.

\para{Elimination of heavy vertices.}
In the procedure of cutting  balls, it is possible that the ball itself contains nearly all the vertices. In this case, the recursion depth can be large, affecting both the cut probability and running time. Our second contribution is a modular reduction to the special case where all out-balls and in-balls contain at most 75\% of the edges. We remark that the threshold of 75\% is arbitrary and can be replaced by any constant, even less than 50\%. We expect this reduction to be useful in future developments, since it reduces the general problem to a setting that is easier to reason about.

To describe this reduction, call a vertex out-heavy if its out-ball contains more than half the edges, and in-heavy if its in-ball contains more than half the edges. The algorithm can detect such vertices by computing single-source shortest paths from $O(\log n)$ many random sources~\cite{bringmann2023negative}. There are now two cases.

First, if there is an in-heavy vertex $s$ and an out-heavy vertex $t$ such that $d(s,t)\lesssim\De$ (which includes the case $s=t$), then we show that the intersection of the in-ball $B^-_G(s,r)$ and out-ball $B^+_G(t,r)$ has weak diameter at most $\De$. Here, the radius $r$ is approximately $\De$ but is random enough to limit the probability that any given edge is cut. The algorithm removes edges cut by these balls, and recursively solves all components except $B^-_G(s,r)\cap B^+_G(t,r)$.

Otherwise, all in-heavy vertices are far from out-heavy vertices (in the direction from in-heavy to out-heavy), and the algorithm computes a random cut between in-heavy and out-heavy vertices. Each side either contains only in-heavy vertices or contains only out-heavy vertices, and sampling enough out-balls and in-balls, respectively, eliminates these vertices with high probability.

\section{Main Algorithm}\secl{1}

Call a vertex \emph{out-light} if $|E[B^+_G(v,\De/8)]|\le\f34m$, and \emph{in-light} if $|E[B^-_G(v,\De/8)]|\le\f34m$. In this section, we assume for simplicity that all vertices are out-light and in-light, though we explicitly mention whenever this assumption is needed. We develop a modular reduction to this assumption in \sec{2}.

At a high level, the algorithm is recursive, cutting out-balls and in-balls of a random radius for a number of iterations. The algorithm is simple enough to state completely, so we present it in pseudocode below.
\begin{algo} \label{algo:1} Let $G=(V,E,w)$ be the input graph on $n$ vertices and $m$ edges, and let $\De$ be the diameter parameter.
\BE
\im Initialize $U=V$. Throughout the algorithm, $U$ is the set of remaining vertices.\label{step:1}
\im Let $L=\lc\lg\lg m\rc$, and define the decreasing sequence $a_0=\De/8$ and $a_i=a_{i-1}-\f\De{16\min\{L,\,2^i\}}$ for $i\in\{1,2,\lds,L\}$.
\im For $i\in\{1,2,\lds,L\}$ in increasing order:
 \BE
 \im Sample each vertex in $U$ with probability $\min\{1,\,4\cd 2^{2^i}\ln(m\De)\cd\f{\deg(v)}{2m}\}$. Let $S_i$ be the random sample.\label{step:a}
 \im Sample a random radius $r_i\in[a_i,a_{i-1}]$.
 \im For each vertex $v\in S_i$ in a random order:\label{step:c}
  \BE
  \im Cut the out-ball $B^+_G(v,r_i)$. That is, recursively solve $G[B^+_G(v,r_i)\cap U]$, and then set\linebreak $U\gets U\sm B^+_G(v,r_i)$.
  \im Cut the in-ball $B^-_G(v,r_i)$. That is, recursively solve $G[B^-_G(v,r_i)\cap U]$, and then set\linebreak $U\gets U\sm B^-_G(v,r_i)$.
  \EE
 \EE
\EE
\end{algo}
\begin{remark}
The sequence $a_0,a_1,\lds,a_L$ is non-negative since
\[ \sum_{i=1}^L\f\De{16\min\{L,\,2^i\}}\le\sum_{i=1}^L\lp\f\De{16L}+\f\De{16\cd2^i}\rp=\f1{16}\bigg(\sum_{i=1}^L\f\De{L}+\sum_{i=1}^L\f\De{2^i}\bigg)\le\f1{16}(\De+\De)=\f\De8=a_0 ,\]
so the algorithm is computing balls of non-negative radius.
\end{remark}
\begin{remark}
The algorithm has no base cases, since those may include vertices that are not out-light or in-light. In other words, the assumption that all vertices are out-light and in-light is not fully general. We handle this issue in \sec{2}.
\end{remark}

\BL\leml{2}
At the beginning of each iteration $i>1$, with probability at least $1-2(m\De)^{-3}$, each vertex $u\in U$ satisfies $\vol(B^\pm_G(u,a_{i-1})\cap U)\le 2m/2^{2^{i-1}}$.
\EL
\BP
For simplicity, we only consider the case of in-balls.  Consider a vertex $u\in U$ with $\vol(B^-_G(u,a_{i-1})\cap U)\ge 2m/2^{2^{i-1}}$ at the beginning of iteration $i-1$. Our goal is to show that $u$ is almost certainly removed from $U$ on iteration $i-1$. On that iteration, each vertex in $U$ is sampled with probability $\min\{1,\,4\cd 2^{2^{i-1}}\ln(m\De)\cd\f{\deg(v)}{2m}\}$, so the probability of sampling no vertex in $B^-_G(u,a_{i-1})\cap U$ is
\begin{align*}
&\prod_{v\in B^-_G(u,a_{i-1})\cap U}\lp1-\min\left\{1,\,4\cd 2^{2^{i-1}}\ln(m\De)\cd\f{\deg(v)}{2m}\right\}\rp
\\\le{}&\prod_{v\in B^-_G(u,a_{i-1})\cap U}\exp\lp-4\cd 2^{2^{i-1}}\ln(m\De)\cd\f{\deg(v)}{2m}\rp
\\={}&\exp\lp-4\cd 2^{2^{i-1}}\ln(m\De)\cd\f{\vol(B^-_G(u,a_{i-1})\cap U)}{2m}\rp
\\\le{}&\exp\bigg(-4\cd 2^{2^{i-1}}\ln(m\De)\cd\f{2m/2^{2^{i-1}}}{2m}\bigg)
\\={}&\exp(-4\ln(m\De))
\\={}&(m\De)^{-4}.
\end{align*}
That is, with probability at least $1-(m\De)^{-4}$, at least one vertex $v\in B^-_G(u,a_{i-1})$ is sampled on iteration $i-1$. In this case, since $r_{i-1}\ge a_{i-1}$, we also have $v\in B^-_G(u,r_{i-1})$, or equivalently $u\in B^+_G(v,r_{i-1})$, so vertex $u$ is removed from $U$ when cutting $B^+_G(v,r_{i-1})$ (if not earlier). By a union bound over all $u\in U$, with probability at least $1-(m\De)^{-3}$ the same holds for all $u\in U$ with $\vol(B^-_G(u,a_{i-1})\cap U)\ge 2m/2^{2^{i-1}}$ at the beginning of iteration $i-1$. In this case, the only vertices that remain in $U$ after iteration $i-1$ must have $\vol(B^-_G(u,a_{i-1})\cap U)\le 2m/2^{2^{i-1}}$ for the set $U$ at the beginning of iteration $i-1$. Since the set $U$ can only shrink over time, the same bound holds at the beginning of iteration $i$, as promised.

The proof for out-balls $\vol(B^+_G(u,a_{i-1})\cap U)\le 2m/2^{2^{i-1}}$ is symmetric, and we union bound over both cases for a final probability of $1-2(m\De)^{-3}$.
\EP
\BL\leml{4}
Consider an edge $(u,v)$. For each $1\le i\le L$, let $p_i$ be the probability that $u,v\in U$ at the beginning of iteration $i$. Then, edge $(u,v)$ is cut on iteration $i$ with probability at most $p_i\cd O(\f{2^i\log\log(m\De)}{\De})\cd w(u,v)$.
\EL
\BP
Assume that $u,v\in U$ at the beginning of iteration $i$, which holds with probability $p_i$ by definition. It suffices to show that the probability of cutting edge $(u,v)$ on iteration $i$, conditioned on this assumption, is $O(\f{2^i\log\log(m\De)}{\De})\cd w(u,v)$.

For simplicity, we only consider out-balls $B^+_G(t,r_i)$.  Let random variable $k$ be the number of sampled vertices in $U$ at the beginning of iteration $i$. The expected value of $k$ is
\begin{align*}
\sum_{v\in U}\min\left\{1,\,4\cd 2^{2^i}\ln(m\De)\cd\f{\deg(v)}{2m}\right\}&\le4\cd 2^{2^i}\ln(m\De)\cd\f{\vol( U)}{2m}
\\&\le4\cd 2^{2^i}\ln(m\De)\cd\f{\vol(V)}{2m}
\\&=4\cd 2^{2^i}\ln(m\De). 
\end{align*}

Given the value of $k$, order the sampled vertices $t_1,\lds,t_k$ by value of $d(t_j,u)$, with ties broken arbitrarily. Recall that the algorithm iterates over $t_1,\lds,t_k$ in a random order. In order for ball $B^+_G(t_j,r_i)$ to cut edge $(u,v)$, two conditions must hold:
 \BE
 \im The radius $r_i$ must satisfy $d(t_j,u)\le r_i<d(t_j,u)+w(u,v)$, since this holds whenever $u\in B^+_G(t_j,r_i)$ and $v\notin B^+_G(t_j,r_i)$. There are $a_{i-1}-a_i=\f\De{16\min\{L,\,2^i\}}$ many choices of $r_i$, so this happens with probability at most $\f{16\min\{L,\,2^i\}}{\De}\cd w(u,v)$.
 \im The vertex $t_j$ must precede $t_1,\lds,t_{j-1}$ in the ordering, since otherwise another ball would claim $u$ first. Since the ordering is random, this happens with probability $1/j$.
 \EE
The two events above are independent, so the probability that ball $B^+_G(t_j,r_i)$ cuts edge $(u,v)$ is at most $\f{16\min\{L,\,2^i\}}\De\cd w(u,v)\cd\f1j$. Summing over all $j$, the overall probability that edge $(u,v)$ is cut by a ball $B^+_G(t_j,r_i)$ is at most
\[ \f{16\min\{L,\,2^i\}}\De\cd\lp\f11+\f12+\cds+\f1k\rp\cd w(u,v)\le \f{16\min\{L,\,2^i\}}\De\cd(\ln k+1)\cd w(u,v) .\]
Taking the expectation over $k$, and using the fact that $\ln k$ is concave, the overall probability is at most
\begin{align*}
&\f{16\min\{L,\,2^i\}}\De\cd(\E[\ln k]+1)\cd w(u,v)
\\\le{}&\f{16\min\{L,\,2^i\}}\De\cd(\ln\E[k]+1)\cd w(u,v)
\\\le{}&\f{16\min\{L,\,2^i\}}\De\cd(\ln(4\cd2^{2^i}\ln(m\De))+1)\cd w(u,v)
\\\le{}&O\lp\f{\min\{\log\log(m\De),\,2^i\}\cd(2^i+\log\log(m\De))}{\De}\rp\cd w(u,v)
\\\le{}&O\lp\f{2^i\log\log(m\De)}\De\rp\cd w(u,v) ,
\end{align*}
where the last step uses the inequality $\min\{a,b\}\cd(a+b)\le\min\{a,b\}\cd2\max\{a,b\}=2ab$.

Finally, the analysis is symmetric for in-balls, so the overall cut probability at most doubles.
\EP

\BL\leml{5}
Assume that each vertex is always out-light and in-light in every recursive instance. At the end of the recursive algorithm, each edge $(u,v)$ is cut with probability $O(\f{\log m\log\log(m\De)}\De)\cd w(u,v)$.
\EL
\BP
Consider a function $f(m)$ such that any edge $(u,v)$ is cut with probability at most $f(m)\cd w(u,v)$ on any graph with at most $m$ edges. Consider an instance of the recursive algorithm. On each iteration $i$, by \lem{4}, a fixed edge $(u,v)$ is cut with probability at most $p_i\cd\f{C\cd2^i\log\log(m\De)}{\De}\cd w(u,v)$ for some constant $C>0$. By the definition of $p_i$, the probability that $u,v\in U$ at the beginning of iteration $i$ but not at the end of iteration $i$ is exactly $p_i-p_{i+1}$, where we define $p_{L+1}=0$. (Note that all vertices leave $U$ after iteration $L$ since every vertex is sampled on that iteration.) In particular, with probability at most $p_i-p_{i+1}$, both $u$ and $v$ are cut out by some common ball $B^\pm_G$ on iteration $i$, which opens the possibility that the edge is cut recursively. If $i=1$, then the recursive instance $G[B^\pm_G\cap U]$ has at most $\f34m=1.5m/2^{2^{i-1}}$ edges since all vertices are out-light and in-light by assumption. If $i>1$, then by adding an overall $2(m\De)^{-3}$ to the cut probability, we may assume that \lem{2} holds on iteration $i$, so $|E[B^\pm_G\cap U]|\le\vol(B^\pm_G\cap U)/2\le m/2^{2^{i-1}}\le1.5m/2^{2^{i-1}}$. Summing over all $i$, we can bound the cut probability divided by $w(u,v)$ by
\begin{align*}
&\sum_{i=1}^L\lp p_i\cd\f{C\cd2^i\log\log(m\De)}{\De}+(p_i-p_{i+1})f(1.5m/2^{2^{i-1}})\rp+2L(m\De)^{-3}
\\={}&\sum_{i=1}^L\lp\big((p_i-p_{i+1})+(p_{i+1}-p_{i+2})+\cds+(p_L-p_{L+1})\big)\cd\f{C\cd2^i\log\log(m\De)}{\De}+(p_i-p_{i+1})f(1.5m/2^{2^{i-1}})\rp
\\&\hspace{13.8cm}+2L(m\De)^{-3}
\\={}&\sum_{i=1}^L(p_i-p_{i+1})\cd\lp\f{C\cd(2^1+2^2+\cds+2^i)\log\log(m\De)}{\De}+f(1.5m/2^{2^{i-1}})\rp+2L(m\De)^{-3}
\\\le{}&\sum_{i=1}^L(p_i-p_{i+1})\cd\lp\f{C\cd2^{i+1}\log\log(m\De)}{\De}+f(1.5m/2^{2^{i-1}})\rp+2L(m\De)^{-3}.
\end{align*}
Since the values $(p_i-p_{i+1})$ sum to $1$, the summation is a convex combination of the terms $\big(\f{C\cd2^{i+1}\log\log(m\De)}{\De}+f(1.5m/2^{2^{i-1}})\big)$, so we can bound the summation by the maximum such term. In conclusion, we have the recursive bound
\[ f(m)\le\max_{i=1}^L\lp\f{C\cd2^{i+1}\log\log(m\De)}{\De}+f(1.5m/2^{2^{i-1}})\rp+2L(m\De)^{-3} .\]
We now solve the recurrence by $f(m)\le\f{18C\lg m\log\log(m\De)}\De$. By induction on $m$, we bound each term in the maximum above by
\begin{align*}
\f{C\cd2^{i+1}\log\log(m\De)}{\De}+f(1.5m/2^{2^{i-1}})&\le\f{C\cd2^{i+1}\log\log(m\De)}{\De}+\f{18C\lg(1.5m/2^{2^{i-1}})\log\log(m\De)}\De
\\&=\f{2C\cd2^i\log\log(m\De)}{\De}+\f{18C(\lg m+\lg 1.5-2^{i-1})\log\log(m\De)}\De
\\&\le\f{2C\cd2^i\log\log(m\De)}{\De}+\f{18C(\lg m-\f13\cd2^{i-1})\log\log(m\De)}\De
\\&=\f{18C\lg m\log\log(m\De)}\De-\f{C\cd2^i\log\log(m\De)}\De,
\end{align*}
and therefore
\begin{align*}
f(m)&\le\max_{i=1}^L\lp\f{C\cd2^{i+1}\log\log(m\De)}{\De}+f(1.5m/2^{2^{i-1}})\rp+2L(m\De)^{-3}
\\&\le\max_{i=1}^L\lp\f{18C\lg m\log\log(m\De)}\De-\f{C\cd2^i\log\log(m\De)}\De\rp+2L(m\De)^{-3}
\\&\le\f{18C\lg m\log\log(m\De)}\De ,
\end{align*}
completing the induction.
\EP

\BL\leml{6}
The algorithm can be implemented in $O((m+n\log\log n)\log^2n)$ time with high probability on graphs with integral and polynomially-bounded edge weights.
\EL
\BP
Consider the algorithm on a given recursive instance. The most expensive step is computing the sets $B^\pm_G(v,r_i)\cap U$, which requires single-source shortest paths computations. For simplicity, we only consider out-balls $B^+_G(v,r_i)\cap U$, since the other case is symmetric. To implement this step, we run Dijkstra's algorithm for each $v\in S_i$ in the random ordering with one important speedup. Suppose that we are currently computing $B^+_G(v,r_i)\cap U$ for some $v\in S_i$ and are currently processing vertex $u\in B^+_G(v,r_i)$ popped from the priority queue in Dijkstra's algorithm. If there is a vertex $v'\in S_i$ that precedes $v$ in the random ordering such that $d(v',u)\le d(v,u)$, then skip $u$ and continue to the next vertex popped from the priority queue. That is, we do not add the neighbors of $u$ to the priority queue. We now justify why this speedup does not affect our goal of computing $B^+_G(v,r_i)\cap U$. If the algorithm skips $u$, then for any vertex $t\in B^+_G(v,r_i)$ whose shortest path from $v$ passes through $u$, there is a shorter path from $v'$ to $t$ (also passing through $u$), so we must have $t\in B^+_G(v',r_i)$ as well. Since $v'$ precedes $v$ in the random ordering, the vertex $t$ is already removed from $U$ when $v'$ is processed (if not earlier). Hence, our speedup only skips over vertices that no longer belong to $U$.

We now bound the number of times a vertex $u$ can be added to Dijkstra's algorithm over all computations of $B^+_G(v,r_i)\cap U$. Recall from the proof of \lem{4} that the number $k$ of sampled vertices in $U$ is at most $4\cd 2^{2^i}\ln(m\De)$. Order the sampled vertices $v_1,\lds,v_k$ by value of $d(v_j,u)$, with ties broken arbitrarily. For each vertex $v_j$, if any of $v_1,\lds,v_{j-1}$ precedes $v_j$ in the random ordering, then Dijkstra's algorithm from $v_j$ skips $u$. It follows that vertex $u$ is added to the priority queue of Dijkstra's algorithm from $v_j$ with probability at most $1/j$. The expected number of times vertex $u$ is added to a priority queue is at most $\f11+\f12+\cds+\f1k\le\ln k+1$. Taking the expectation over $k$, and using the fact that $\ln k$ is concave, this expectation is at most $\ln(4\cd 2^{2^i}\ln(m\De))+1=O(2^i+\log\log(m\De))$. Summing over all $L=\lc\lg\lg m\rc$ iterations, the expected total number of times vertex $u$ is added to a priority queue is $O(2^1+2^2+\cds+2^L+L\cd\log\log(m\De))=O(\log n)$, where we assume that $\De=\poly(n)$ since edge weights are polynomially bounded. Using Thorup's priority queue~\cite{thorup2003integer}, each vertex update takes $O(\log\log n)$ time and each edge update takes $O(1)$ time, so the expected total running time is $O((m+n\log\log n)\log n)$. Finally, since the algorithm has recursion depth $O(\log n)$, the overall expected running time picks up another $O(\log n)$ factor. The running time can even occur with high probability by an argument from~\cite{bringmann2023negative}: restart each recursive instance (before making any recursive calls) if the running time of that instance exceeds twice its expectation. By Markov's inequality, each attempt succeeds with probability at least $1/2$, so edge cut probabilities at most double since we are conditioning on an event that occurs with probability at least $1/2$. By a Chernoff bound, each vertex and edge participates in $O(\log n)$ many restarts over the $O(\log n)$ levels of recursion with high probability.
\EP

\section{Eliminating Heavy Vertices}\secl{2}

Recall that a vertex is \emph{out-light} if $|E[B^+_G(v,\De/8)]|\le\f34m$, and \emph{in-light} if $|E[B^-_G(v,\De/8)]|\le\f34m$. Define a vertex as \emph{out-heavy} if $|E[B^+_G(v,\De/8)]|\ge\f12m$,  and \emph{in-heavy} if $|E[B^-_G(v,\De/8)]|\ge\f12m$. Note that a vertex can be both out-light and out-heavy, and both in-light and in-heavy. We remark that the thresholds for light and heavy vertices are arbitrary and can be replaced by $\al m$ and $\be m$ for any constants $0<\be<\al<1$. We stick to $\f34m$ and $\f12m$ since these are the thresholds defined in~\cite{bringmann2023negative}, from which we use the following heavy-light classification algorithm.
\BL[Lemma~19 of~\cite{bringmann2023negative}]
There is an algorithm using $O(\log n)$ single-source shortest paths computations that, with high probability, correctly labels each vertex as either out-light or out-heavy, and either in-light or in-heavy. Vertices that are both out-light and out-heavy may receive either label, and the same holds for both in-light and in-heavy.
\EL
From now on, we re-define out-light, out-heavy, in-light, and in-heavy based on the labels of the above algorithm, which we assume to be correct. In particular, vertices that were both out-light and out-heavy are now only one and not the other, depending on which label was assigned by the algorithm. Similarly, vertices that were both in-light and in-heavy are now one and not the other.  There are now two cases.
\para{Case 1.} Suppose that there exists in-heavy vertex $s$ and out-heavy vertex $t$ such that $d(s,t)\le\De/4$. The algorithm can find such vertices (if they exist) by contracting all in-heavy vertices into a vertex $s'$, contracting all out-heavy vertices into a vertex $t'$, and computing the distance from $s'$ to $t'$, which measures the shortest distance from any in-heavy vertex to any out-heavy vertex.

In this case, the algorithm samples a random radius $r\in[\De/8,\De/4]$ and performs the following:
  \BE
  \im[i.] Cut the in-ball $B^-_G(s,r)$ and recursively solve the complement $G[V\sm B^-_G(s,r)]$.
  \im[ii.] Cut the out-ball $B^+_G(t,r)$ and recursively solve the (remaining) complement $G[B^-_G(s,r)\sm B^+_G(t,r)]$.
  \EE
The two recursive problems have at most $\f12m$ edges each since they do not contain edges in $B^-_G(s,r)$ and $B^+_G(t,r)$, respectively. Now consider the remaining set $B^-_G(s,r)\cap B^+_G(t,r)$. The lemma below implies that the algorithm does not need to recursively solve $G[B^-_G(s,r)\cap B^+_G(t,r)]$.
\BL
$B^-_G(s,r)\cap B^+_G(t,r)$ has weak diameter at most $\De$.
\EL
\BP
Consider two vertices $u,v\in B^-_G(s,r)\cap B^+_G(t,r)$. Since $u\in B^-_G(s,r)$, we have $d(u,s)\le r$, and since $v\in B^+_G(t,r)$, we have $d(t,v)\le r$. Also, we have $d(s,t)\le\De/4$ by the assumption of Case~1. Altogether, we have $d(u,v)\le d(u,s)+d(s,t)+d(t,v)\le r+\De/4+r\le\De$.
\EP
We now bound the probability of cutting each edge. In order for ball $B^+_G(t,r)$ to cut edge $(u,v)$, the radius $r$ must satisfy $d(t,u)\le r<d(t,u)+w(u,v)$, since this holds whenever $u\in B^+_G(t,r)$ and $v\notin B^+_G(t,r)$. There are $\De/8$ many choices of $r$, so this happens with probability at most $\f8\De\cd w(u,v)$. Similarly, the edge $(u,v)$ is cut by ball $B^-_G(s,r)$ with probability at most $\f8\De\cd w(u,v)$. Overall, each edge $(u,v)$ is cut with probability at most $\f{16}\De\cd w(u,v)$.

\para{Case 2.} Suppose that the assumption of Case~1 does not hold. Sample a random radius $r\in[\De/8,\De/4]$, and compute the union $B^+_G$ of all balls $B^+_G(s,r)$ where $s$ is in-heavy. By assumption, there are no out-heavy vertices in $B^+_G$. Also, by construction, all in-heavy vertices are in $B^+_G$. The algorithm can find $B^+_G$ by contracting all in-heavy vertices into a vertex $s'$ and computing $B^+_G(s',r)$ in the contracted graph.

The algorithm cuts $B^+_G$ and recursively solves whichever of $G[B^+_G]$ and $G[V\sm B^+_G]$ has fewer edges. Each edge $(u,v)$ is cut with probability at most $\f8\De\cd w(u,v)$ by a similar analysis. Suppose first that the algorithm recursively solves $G[B^+_G]$ with at most $\f12m$ edges and leaves $V\sm B^+_G$ which has no in-heavy vertices. The algorithm eliminates all out-heavy vertices by executing the following, which is essentially steps~(\ref{step:a}) to~(\ref{step:c}) of \Cref{algo:1} for iteration $i=1$, except that only in-balls are cut (and the random radius has a slightly different range).
\begin{algo}\label{algo:2} \
 \BE
 \im[1.] Initialize $U=V\sm B^+_G$. Throughout the algorithm, $U$ is the set of remaining vertices.
 \im[(3a)] Sample each vertex in $U$ with probability $O(\ln(m\De)\cd\f{\deg(v)}{2m})$. Let $S_1$ be the random sample.
 \im[(3b)] Sample a random radius $r_1\in[\f\De8,\f\De4]$.
 \im[(3c)] For each vertex $v\in S_1$ in a random order:
  \BE
  \im[i.] Cut the in-ball $B^-_G(v,r_1)$. That is, recursively solve $G[B^-_G(v,r_1)\cap U]$, and then set\linebreak $U\gets U\sm B^-_G(v,r_1)$.
  \EE
 \EE
\end{algo}
Since there are no in-heavy vertices in $V\sm B^+_G$, each recursive call has at most $\f12m$ edges. Each edge is cut with probability $O(\f{\log\log(m\De)}{\De})\cd w(u,v)$, which can be analyzed identically to the proof of \lem{4} (for iteration $i=1$).
\BL
With high probability, the final set $U$ has no out-heavy vertices.
\EL
\BP
Consider an out-heavy vertex $u$, which means that $|E[B^+_G(u,\De/8)]|\ge\f12m$. By a similar calculation to the proof of \lem{2}, with high probability at least one vertex $v\in B^+_G(u,\De/8)$ is sampled. In this case, since $r_1\ge\De/8$, we also have $v\in B^+_G(u,r_1)$, or equivalently $u\in B^-_G(v,r_1)$, so vertex $u$ is removed from $U$ when cutting out $B^-_G(v,r_1)$ (if not earlier). Taking a union bound over all out-heavy vertices, we conclude that with high probability, the final set $U$ has no out-heavy vertices.
\EP
Since there are no in-heavy vertices in the initial set $V\sm B^+_G$, we conclude that all vertices in $U$ are both out-light and in-light. Finally, the case when the algorithm cuts out $V\sm B^+_G$ is analogous, with out-balls replaced by in-balls and vice versa.

\para{Combining both cases.}
In both cases, the algorithm recursively solves instances of at most $\f12m$ edges. In Case~1, the algorithm declares that the remaining component has weak diameter at most $\De$, and \Cref{algo:1} is skipped. In Case~2, the algorithm executes \Cref{algo:1} starting with the set $U$ at the end of \Cref{algo:2} (instead of initializing it to $V$ on step~\ref{step:1}). Since all vertices in $U$ are both out-light and in-light, the guarantees of \Cref{algo:1} apply.

We conclude with the overall running time and edge cut probability, establishing \thm{main}.
\BL
With the heavy vertex removal process added to each instance of \Cref{algo:1}, the algorithm can be implemented in $O((m+n\log\log n)\log^2n)$ time with high probability on graphs with integral and polynomially-bounded edge weights.
\EL
\BP
The heavy vertex removal process uses $O(\log n)$ single-source shortest paths computations, which can be implemented in $O((m+n\log\log n)\log n)$ time using Thorup's priority queue. The recursion tree has depth $O(\log n)$, so this is $O((m+n\log\log n)\log^2n)$ additional time, which is subsumed by the running time from \lem{6}.
\EP
\BL
With the heavy vertex removal process added to each instance of \Cref{algo:1}, each edge $(u,v)$ is cut with probability $O(\f{\log n\log\log n}\De)\cd w(u,v)$.
\EL
\BP
In both cases, each edge is cut with probability $O(\f{\log\log(m\De)}\De)\cd w(u,v)$. The recursion tree has depth $O(\log m)$, so each edge has an additional $O(\f{\log m\log\log(m\De)}{\De})\cd w(u,v)$ probability of being cut by the heavy vertex removal process, which is subsumed by the probability bound from \lem{5}. Overall, each edge $(u,v)$ is cut with probability $O(\f{\log m\log\log(m\De)}\De)\cd w(u,v)=O(\f{\log n\log\log n}\De)\cd w(u,v)$, where we assume that $\De=\poly(n)$ since edge weights are polynomially bounded.
\EP

\section*{Acknowledgement}
We thank Satish Rao and Thatchaphol Saranurak for helpful references on the history of directed LDDs.

\bibliographystyle{alpha}
\bibliography{ref}

\newcommand{\etalchar}[1]{$^{#1}$}
\begin{thebibliography}{BNWN25}

\bibitem[ABCP92]{awerbuch1992fast}
Baruch Awerbuch, Bonnie Berger, Lenore Cowen, and David Peleg.
\newblock Fast network decomposition.
\newblock In {\em Proceedings of the eleventh annual ACM symposium on
  Principles of distributed computing}, pages 169--177, 1992.

\bibitem[AGLP89]{awerbuch1989network}
Baruch Awerbuch, Andrew~V Goldberg, Michael Luby, and Serge~A Plotkin.
\newblock Network decomposition and locality in distributed computation.
\newblock In {\em FOCS}, volume~30, pages 364--369. Citeseer, 1989.

\bibitem[Bar96]{bartal1996probabilistic}
Yair Bartal.
\newblock Probabilistic approximation of metric spaces and its algorithmic
  applications.
\newblock In {\em Proceedings of 37th Conference on Foundations of Computer
  Science}, pages 184--193. IEEE, 1996.

\bibitem[BCF23]{bringmann2023negative}
Karl Bringmann, Alejandro Cassis, and Nick Fischer.
\newblock Negative-weight single-source shortest paths in near-linear time: Now
  faster!
\newblock In {\em 2023 IEEE 64th Annual Symposium on Foundations of Computer
  Science (FOCS)}, pages 515--538. IEEE, 2023.

\bibitem[BFHL25]{bringmann2025near}
Karl Bringmann, Nick Fischer, Bernhard Haeupler, and Rustam Latypov.
\newblock Near-optimal directed low-diameter decompositions.
\newblock In {\em International Colloquium on Automata, Languages, and
  Programming (ICALP)}, 2025.

\bibitem[BGK{\etalchar{+}}11]{blelloch2011near}
Guy~E Blelloch, Anupam Gupta, Ioannis Koutis, Gary~L Miller, Richard Peng, and
  Kanat Tangwongsan.
\newblock Near linear-work parallel sdd solvers, low-diameter decomposition,
  and low-stretch subgraphs.
\newblock In {\em Proceedings of the twenty-third annual ACM symposium on
  Parallelism in algorithms and architectures}, pages 13--22, 2011.

\bibitem[BNWN25]{bernstein2025negative}
Aaron Bernstein, Danupon Nanongkai, and Christian Wulff-Nilsen.
\newblock Negative-weight single-source shortest paths in near-linear time.
\newblock {\em Communications of the ACM}, 68(2):87--94, 2025.

\bibitem[CKR00]{cualinescu2000improved}
Gruia C{\u{a}}linescu, Howard Karloff, and Yuval Rabani.
\newblock An improved approximation algorithm for multiway cut.
\newblock {\em Journal of Computer and System Sciences}, 60(3):564--574, 2000.

\bibitem[CZ20]{chechik2020dynamic}
Shiri Chechik and Tianyi Zhang.
\newblock Dynamic low-stretch spanning trees in subpolynomial time.
\newblock In {\em Proceedings of the Fourteenth Annual ACM-SIAM Symposium on
  Discrete Algorithms}, pages 463--475. SIAM, 2020.

\bibitem[FG19]{forster2019dynamic}
Sebastian Forster and Gramoz Goranci.
\newblock Dynamic low-stretch trees via dynamic low-diameter decompositions.
\newblock In {\em Proceedings of the 51st Annual ACM SIGACT Symposium on Theory
  of Computing}, pages 377--388, 2019.

\bibitem[KPRT97]{klein1997approximation}
Philip~N Klein, Serge~A Plotkin, Satish Rao, and Eva Tardos.
\newblock Approximation algorithms for steiner and directed multicuts.
\newblock {\em Journal of Algorithms}, 22(2):241--269, 1997.

\bibitem[LR99]{leighton1999multicommodity}
Tom Leighton and Satish Rao.
\newblock Multicommodity max-flow min-cut theorems and their use in designing
  approximation algorithms.
\newblock {\em Journal of the ACM (JACM)}, 46(6):787--832, 1999.

\bibitem[LS93]{linial1993low}
Nathan Linial and Michael Saks.
\newblock Low diameter graph decompositions.
\newblock {\em Combinatorica}, 13(4):441--454, 1993.

\bibitem[MPX13]{miller2013parallel}
Gary~L Miller, Richard Peng, and Shen~Chen Xu.
\newblock Parallel graph decompositions using random shifts.
\newblock In {\em Proceedings of the twenty-fifth annual ACM symposium on
  Parallelism in algorithms and architectures}, pages 196--203, 2013.

\bibitem[Sey95]{seymour1995packing}
Paul~D. Seymour.
\newblock Packing directed circuits fractionally.
\newblock {\em Combinatorica}, 15(2):281--288, 1995.

\bibitem[Tho03]{thorup2003integer}
Mikkel Thorup.
\newblock Integer priority queues with decrease key in constant time and the
  single source shortest paths problem.
\newblock In {\em Proceedings of the thirty-fifth annual ACM symposium on
  Theory of computing}, pages 149--158, 2003.

\end{thebibliography}

\end{document}